\documentclass[10pt, journal]{IEEEtran} 

\usepackage{amsmath,amssymb,algorithm,algorithmic,cite}
\usepackage{tikz}
\usetikzlibrary{positioning,calc,fit,shapes,arrows,backgrounds,plotmarks,arrows.meta}
\usepackage{subfig}
\usepackage{pgfplots}
\usetikzlibrary{external} 
\usepackage{grffile}
\usepackage[mode=buildnew]{standalone}
\usepackage[keeplastbox]{flushend}
\usepackage{multirow}
\usepackage{psfrag}
\usepackage{graphicx}
\usepackage{xcolor,colortbl}
\usepackage{hhline}
\usetikzlibrary{external}
\usepackage{acronym}
\usepackage{lipsum}
\usepackage{hyperref}
\usepackage{booktabs}
\usepackage{footnote}
\makesavenoteenv{tabular}
\makesavenoteenv{table}
\usepackage{multirow}

\acrodef{MVDR}[MVDR]{Minimum Variance Distortionless Response}
\acrodef{GMVDR}[GMVDR]{Generalized MVDR}
\acrodef{GSC}[GSC]{Generalized Sidelobe Canceller}
\acrodef{GEV}[GEV]{Generalized Eigenvalue}
\acrodef{MWF}[MWF]{Multichannel Wiener Filter}
\acrodef{MMSE}[MMSE]{Minimum Mean Square Error}
\acrodef{DNN}[DNN]{Deep Neural Network}
\acrodef{MNMF}[MNMF]{Multichannel Non-negative Matrix Factorization}
\acrodef{LSTM}[LSTM]{Long-Short Term Memory}
\acrodef{STFT}[STFT]{Short Time Fourier Transform}
\acrodef{ATF}[ATF]{Acoustic Transfer Function}
\acrodef{FC}[FC]{Fully Connected}
\acrodef{BN}[BN]{Batch Normalization}
\acrodef{ReLU}[ReLU]{Rectified Linear Unit}
\acrodef{GRU}[GRU]{Gated Recurrent Unit}
\acrodef{SNR}[SNR]{Signal-to-Noise-Ratio}
\acrodef{SDR}[SDR]{Signal-to-Distortion Ratio}
\acrodef{COSPA}[COSPA]{Complex-valued Spatial Autoencoder}
\acrodef{DOA}[DOA]{Direction Of Arrival}
\acrodef{RTF}[RTF]{Relative Transfer Function}
\acrodef{SINR}[SINR]{Signal-to-Interferer and Noise Ratio}
\acrodef{ASR}[ASR]{Automatic Speech Recognition}
\acrodef{RIR}[RIRs]{Room Impulse Responses}
\title{Complex-valued Spatial Autoencoders for Multichannel Speech Enhancement}
 
\author{Mhd Modar Halimeh,~\IEEEmembership{Student member,~IEEE}, and Walter Kellermann,~\IEEEmembership{Fellow,~IEEE} \thanks{M. M. Halimeh and W. Kellermann are with the Chair of Multimedia Communications and Signal
Processing (LMS), University of Erlangen-Nuremberg, 91058 Erlangen, Germany
(e-mail:mhd.m.halimeh@fau.de; walter.kellermann@fau.de).}}

\begin{document}
\maketitle

%%%%%%%%%%%%%%%% ABSTRACT %%%%%%%%%%%%%%%%%%%%%%
\begin{abstract}
In this contribution, we present a novel online approach to multichannel speech enhancement. The proposed method estimates the enhanced signal through a filter-and-sum framework. More specifically, complex-valued masks are estimated by a deep complex-valued neural network, termed the complex-valued spatial autoencoder. The proposed network is capable of exploiting as well as manipulating both the phase and the amplitude of the microphone signals. As shown by the experimental results, the proposed approach is able to exploit both spatial and spectral characteristics of the desired source signal resulting in a physically plausible spatial selectivity and superior speech quality compared to other baseline methods.
\end{abstract}

%%%%%%%%%%%%%%%% Introduction %%%%%%%%%%%%%%%%%%%%%%
\begin{figure*}[t]
    \centering
    \includegraphics[width=0.78\textwidth]{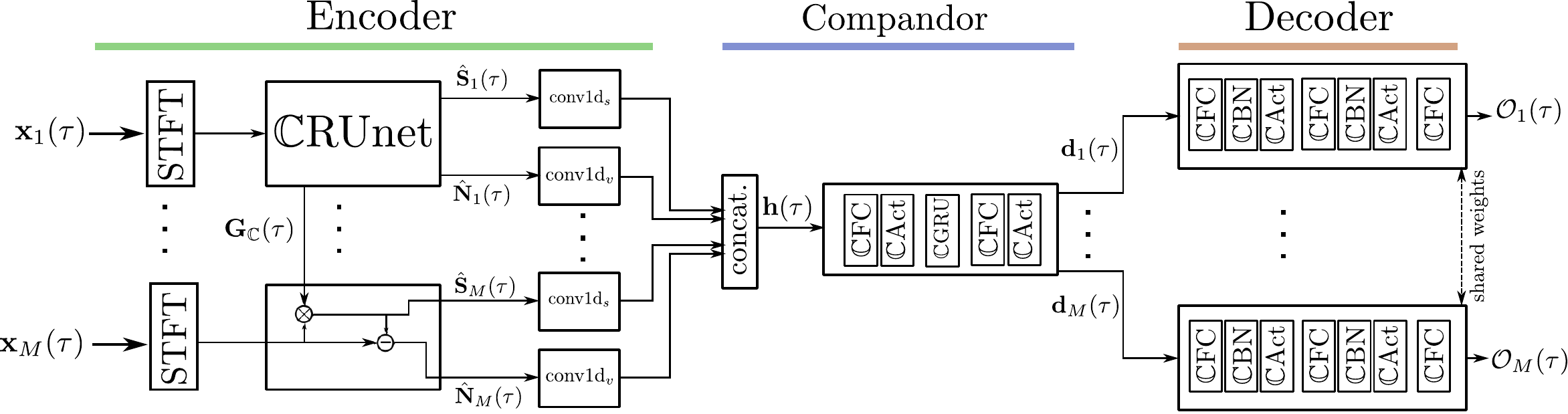}
    \caption{The proposed complex-valued spatial autoencoder structure.}\vspace*{-5mm}
    \label{fig:network}
\end{figure*}

 \vspace*{-3mm} \section{Introduction} \vspace*{-1mm}
        The widespread {availability} of devices with multiple microphones have boosted the interest in multichannel speech enhancement techniques {for}, e.g., source separation, {source extraction, or} noise suppression \cite{Shan85, AdaptiveSIgnalProcessing}.  
        
        {The most commonly used multichannel speech enhancement technique is beamforming, where the spatial diversity of the different sound sources is exploited to emphasize sounds coming from the desired source's direction while suppressing sounds that arrive from other directions \cite{VanVeen,vanTrees, ArrayProcessing}. Many beamformers can be found in the literature derived under different constraints such as the popular \ac{MVDR} beamformer \cite{MicrophoneArraySignalProcessing}, the Generalized MVDR (GMVDR) beamformer \cite{Gannot01}, the \ac{GEV} beamformer \cite{Warsitz07, Pfeifenberger19}, the \ac{MWF} \cite{Doclo02}, and  modulation-domain multichannel Kalman filter \cite{Xue18}.}
        
        In general, conventional beamformers share the need for spatial information, whether in the form of steering vectors or spatial covariance matrices, in order to function properly. Recently, several data-driven methods have been proposed to estimate this information, e.g., in \cite{Donas2020} a combination of a \ac{DNN} {and} a maximum likelihood estimator is used to estimate the clean speech statistics and speech presence probability which are {then} used to compute the beamformer's weights. The authors in \cite{Shimada19} proposed to use the \ac{MNMF} to decompose time-frequency bins into speech and noise components to be used in obtaining the necessary statistics for an \ac{MVDR} beamformer. \ac{MNMF} is replaced by a \ac{DNN}-based speech prior in \cite{Sekiguchi19} to estimate clean speech statistics. 
        
        Alternatively, {departing} from statistically optimum beamformers, a beamformer's weights can be directly estimated using \ac{DNN}s. The authors in \cite{Xiao16} proposed to train a \ac{DNN} to estimate a beamformer's weights {for} maximizing the performance of a subsequent \ac{ASR} system without guaranteeing better speech quality. Similarly, time-domain beamformer weights are estimated using \ac{LSTM} layers in \cite{Li} for better speech recognition performance. Robust speech recognition was also the aim in \cite{Meng17} where {'}deep \ac{LSTM} adaptive beamforming{'} is introduced. Another variant is to infer a time-frequency mask that is applied to the reference microphone to estimate the desired signal. This was done in \cite{Li19} by employing a shared \ac{LSTM} network across subbands and in \cite{Chakrabarty19} using a convolutional recurrent network. Sinc and dilated convolutional layers were used in \cite{Liu20} to perform waveform mapping.   
        
        In this paper, we present a novel approach {to} data-driven online multichannel spatiospectral filtering using complex-valued \ac{DNN}s. The proposed approach adopts the filter-and-sum technique from conventional beamforming as each channel is filtered by a complex-valued mask and the filtered channels are then added to produce the enhanced signal. This allows the network to produce effects such as phase-aligned superposition of the desired signal, in contrast to, e.g., \cite{Li19}. Moreover, unlike, e.g., \cite{Xiao16}, the network is trained for speech quality enhancement and is not a preprocessing block for an \ac{ASR} system, nor is {it} a supporting block for a conventional beamformer as, e.g., in \cite{Donas2020}. Finally, we verify the validity of the proposed approach under various acoustic conditions, where the proposed network is shown to be capable of localizing and extracting the desired speech signal.
        In the following, signals in the \ac{STFT} domain are denoted by uppercase letters while signals in the time domain are denoted by lowercase letters. Furthermore, transposition is denoted by $(\cdot)^\textrm{T}$, '*' denotes conjugate complex, while vectors are denoted by boldface letters.
        
%%%%%%%%%%%%%%%% Spatial Autoencoders %%%%%%%%%%%%%%%%%%%%%%
 \vspace*{-3mm}\section{Complex-valued Spatial Autoencoders}   \vspace*{-1mm}

        We consider a scenario with $M$ microphones, where at time-frequency bin $(\tau,f)$ the $m$-th microphone signal is given by
        \begin{equation} \label{eq:eq_1}
            X_m(\tau, f) = D_m(\tau, f) + N_m(\tau, f),
        \end{equation}
        where $D_m(\tau, f)=H^*_m(\tau, f)S(\tau, f)$ denotes the reverberant source signal{,} $H_m$ denotes the \ac{ATF} from the desired source's position to the $m$-th microphone, while $N_m$ denotes the background noise components as picked up by the $m$-th microphone. It must be pointed out that undesired components captured by $N_m$ are restricted to non-speech components, i.e., interfering speakers are not considered in this work. Nevertheless, $N_m$ is not restricted to stationary nor diffuse noises{,} but it can represent arbitrary noises. Our goal in this paper is to extract the source signal $S$, or a reverberant version of it, with minimal distortions while suppressing the noise components $N$. 

        Complex-valued \ac{DNN}s \cite{trabelsi2018} have shown convincing results in single-channel speech enhancement \cite{Choi2018, Hu2020} as well as in echo suppression \cite{Halimeh_ICASSP2021}. Their ability to manipulate and exploit phase information makes them a natural candidate for our multichannel signal processing task.
        
        The proposed network architecture is shown in Fig.~\ref{fig:network}. As input, the network takes one frame of the time-domain signal $\mathbf{x}_m(\tau)$ comprising $L$ samples per channel $m$, i.e., $\mathbf{x}_m(\tau)=[{x}_m(\tau), ..., {x}_m(\tau-L+1)]^\textrm{T}$ and outputs one complex-valued mask $\boldsymbol{\mathcal{M}}_{m}(\tau)$ per channel. For each time-frequency bin $(\tau, f)$, an estimate of the desired source signal is obtained as follows 
        \begin{equation}  \label{eq:sourceRecon}
            \hat{S}(\tau, f) = \sum_{m=1}^{M} \mathcal{M}_m(\tau, f) \cdot X_m(\tau, f).
        \end{equation}
        
        As seen from the figure, the networks starts by processing each channel's signal separately. Afterwards, information from all channels is processed jointly at the compandor unit in the middle. Finally, each channel's mask is constructed separately. This structure resembles the commonly used autoencoder structures and therefore, we denote it a \textit{spatial autoencoder}. 
%%%%% Encoder Side %%%%% 
 \vspace*{-3mm} \subsection{Spatial Encoders} \vspace*{-1mm}
        As seen in Fig.~\ref{fig:network}, $M$ frames of length $L$ of the $M$ microphone signals $\{\mathbf{x}_m(\tau)\}_{m=1}^{M}$ are processed as follows: First, an \ac{STFT} is performed to obtain $\{\mathbf{X}_m(\tau)\}_{m=1}^{M}$. Afterwards, the complex-valued signal $\mathbf{X}_1(\tau)$ is fed into a complex-valued subnetwork, denoted by $\mathbb{C}$RUnet, that is a smaller variant of the network in \cite{Halimeh_ICASSP2021} consisting of eight complex-valued convolutional modules with a complex-valued \ac{GRU} and a complex-valued \ac{FC} layer in between. The $\mathbb{C}$RUnet produces a complex-valued mask $\mathbf{G}_{\mathbb{C}}(\tau)$ that is used across all $M$ channels to obtain initial estimates of the desired speech components $\mathbf{S}_m(\tau)$ and undesired noise components $\mathbf{N}_m(\tau)$ as
        \begin{align}
            & \hat{\mathbf{S}}_m(\tau) = \mathbf{G}_{\mathbb{C}}(\tau) \odot \mathbf{X}_m(\tau), \\ 
            & \hat{\mathbf{N}}_m(\tau) = (1 - \mathbf{G}_{\mathbb{C}}(\tau)) \odot \mathbf{X}_m(\tau),
        \end{align}
        where $\odot$ denotes the Hadamard product operator. 
        
        The use of the same mask $\mathbf{G}_{\mathbb{C}}(\tau)$ across all microphone signals ensures the preservation of relative phase differences and therefore, the preservation of spatial information as encoded in the original microphone signals. On the other hand, one should acknowledge that using a single complex-valued mask across the different channels {cannot effectuate spatially selective filtering}.

        The initial signal components estimates are then downsampled using two single-dimensional convolutional layers denotes by $\textrm{conv1d}_s$ and $\textrm{conv1d}_n$ to reduce their dimensionality to $L_1<L$. More specifically, the initial source estimates  $\{\hat{\mathbf{S}}_m(\tau)\}_{m=1}^{M}$ are downsampled using $\textrm{conv1d}_s$ that is shared across all $M$ channels, while the noise estimates $\{\hat{\mathbf{N}}_m(\tau)\}_{m=1}^{M}$ are similarly downsampled using $\textrm{conv1d}_n$. This downsampling is done for purely computational purposes as a certain degree of redundancy is to be expected in the signals $\hat{\mathbf{S}}_m(\tau)$ and $\hat{\mathbf{N}}_m(\tau)$.

%%%%% Compandor Side %%%%% 
        \vspace*{-2mm}\subsection{Spatial Compandor}  \vspace*{-1mm} \label{sec:SpatialCompandor}
        The encoders lead to a compandor unit. The goal of the compandor unit is to estimate the necessary complex equalization, or an abstract representation thereof, that adjusts both the amplitude and phase of the different channels in order to extract the desired source {exploiting both the spatial and the spectrotemporal domain}. {As} the compandor is the only part of the network that has access to all channels simultaneously and where different channels {are} processed differently {to lead to the desired spatial selectivity, it is also the part where spatial filtering is accomplished}. Inspired by the coding literature, the term compandor here refers to the compression at the input side, where information from all channels is fused into a single channel stream to be processed jointly, while on the output side the single stream of information is expanded to the original number of channels. More specifically, at the input of the compandor, the signals resulting from the encoding stage are collected in the vector $\mathbf{h}(\tau) \in \mathbb{C}^{2ML_1}$. Therefore, $\mathbf{h}(\tau)$ encapsulates both spatial and spectral information regarding the desired source and any active noise sources.

        The vector $\mathbf{h}(\tau)$ is then processed by a cascade of a complex-valued \ac{FC} layer denoted by ($\mathbb{C}$\ac{FC}), a complex-valued leaky \ac{ReLU} activation function \cite{Choi2018} denoted by ($\mathbb{C}$Act), a complex-valued \ac{GRU} ($\mathbb{C}$\ac{GRU}), and finally a $\mathbb{C}$\ac{FC} and a $\mathbb{C}$Act. These different layers will be characterized by their output sizes which are denoted by $\{L_2, L_3, ML_4\}$, respectively. The inclusion of the the $\mathbb{C}$\ac{GRU} enables the compandor to not only recognize and exploit instantaneous spatial and spectral patterns, but to also exploit the temporal evolution of these patterns.

        Finally, the compandor outputs the vector $\mathbf{d}(\tau) \in \mathbb{C}^{ML_4}$, which is decomposed into $M$ excitation {vectors} $\{\mathbf{d}_m(\tau)\}_{m=1}^{M}$ of length $L_4$ such that each vector is used to construct a complex-valued mask at the decoding stage. 
        
%%%%% Decoder Side %%%%% 
 \vspace*{-3mm} \subsection{Spatial Decoders}  \vspace*{-1mm}\label{sec:SpatialDecoder}   
        Following the compandor unit is the decoding stage, where each excitation vector $\mathbf{d}_m(\tau)$ is fed into a decoder network consisting of a $\mathbb{C}$\ac{FC}, a complex-valued \ac{BN}, and a $\mathbb{C}$Act. This cascade is repeated once more and then followed by the final $\mathbb{C}$\ac{FC} layer. These layers will be characterized by their outputs' dimensions $\{L_5, L_6, L\}$, respectively. The final decoder layer outputs an unprocessed mask $\mathcal{O}_m(\tau, f)$ for each time-frequency bin $(\tau, f)$ that is used to obtain the complex-valued mask $\mathcal{M}_m(\tau, f)$ as follows \cite{Choi2018}
        \begin{equation}
            | {\mathcal{M}}_m(\tau, f)| = \textrm{tanh}(|\mathcal{O}_m(\tau, f)|), 
        \end{equation}
        and 
        \begin{equation}
            e^{i \theta_{\mathcal{M}_m}}(\tau, f) = \frac{\mathcal{O}_m(\tau, f)}{|\mathcal{O}_m(\tau, f)|}.
        \end{equation}
        
        It is worth noting that the aforementioned $M$ decoder networks are identical, i.e., weights are shared across the $M$ decoding channels, and as a consequence, any differences between the $M$ masks $\{{\mathcal{M}}_m(\tau, f) \}_{m=1}^{M}$ can stem only from differences in the excitation vectors $\{\mathbf{d}_m(\tau) \}_{m=1}^{M}$ rather than channel-specific decoder networks. 
        
        Using Eq.~\eqref{eq:sourceRecon} we obtain the \ac{STFT}-domain estimate of the source signal $\hat{\mathbf{S}}(\tau)$ which can be transformed back to time domain to obtain the estimated signal frame $\hat{\mathbf{s}}(\tau)$.

%%%%% Training %%%%% 
 \vspace*{-2mm} \subsection{Training and Optimization}  \vspace*{-1mm}\label{sec:Training}          
        As a training target, we employ the clean reverberant source signals filtered by an \ac{MVDR} beamformer steered towards the source position
         \vspace*{-2mm}\begin{equation} \label{eq:target1}
            S_{\textrm{target}}(\tau, f) = \sum_{m=1}^{M}W^*_m(\tau, f) D_m(\tau, f),
        \end{equation}
        where $W_m(\tau, f)$ denotes an \ac{MVDR} beamformer weight at the $(\tau, f)$ time-frequency bin. The beamformer weights $W_m(\tau, f)$ are calculated using a recursively estimated noise spatial covariance matrix $\mathbf{R}_{NN}(\tau, f)$ using the ground truth noise signals $\{{N}_m(\tau, f) \}_{m=1}^{M}$ and a free-field steering vector towards to the source's ground truth \ac{DOA}. A simple rearrangement of Eq.~\eqref{eq:target1} as a function of the microphone signals yields
        \begin{equation} \label{eq:target2}
            S_{\textrm{target}}(\tau, f) = \sum_{m=1}^{M}W^*_m(\tau, f) \left( cR_{m}(\tau, f) X_m(\tau, f) \right),
        \end{equation}
        where $cR_{m}(\tau, f)$ denotes the ideal complex-valued ratio mask at microphone $m$ and time-frequency bin $(\tau, f)$. Clearly, this {target} is not attainable using only a spatial filter, i.e., $W_m(\tau, f)$, but instead, spectral filtering as represented by $cR_{m}(\tau, f)$ is needed, highlighting the difference to learning a conventional beamformer. Furthermore, compared to utilizing the 'dry' non-reverberant source signal, the proposed target is a reverberant image of the source signal and therefore, dereverberation is not targeted by the network.

        To optimize the network's weights, the \ac{SNR} loss function is used \cite{SNRLoss}
        \begin{equation} 
            \mathcal{J}_{\textrm{SNR}}( \mathbf{s}_{\textrm{target}}(\tau), \hat{\mathbf{s}}(\tau)) = -10 \log_{10}\left( \frac{\Vert \mathbf{s}_{\textrm{target}}(\tau) \Vert^2}{\Vert \mathbf{s}_{\textrm{target}}(\tau)- \hat{\mathbf{s}}(k)\Vert^2} \right), \nonumber
        \end{equation}
        where $\left \Vert \cdot \right \Vert$ denotes the Euclidean norm, while $ \mathbf{s}_{\textrm{target}}(\tau)$ and $\hat{\mathbf{s}}(\tau)$ denote the time-domain target signal and estimated desired signal, respectively. 
        
%%%%%%%%%%%%%%%% Results %%%%%%%%%%%%%%%%%%%%%%
 \vspace*{-3mm} \section{Experimental Results}  \vspace*{-1mm}
For evaluation we compare the proposed approach, denoted by \ac{COSPA} to four different baseline methods: 
\begin{itemize}
    \item The use of the $\mathbb{C}$RUnet as a \ac{DNN}-based single-channel speech enhancement method. This network is trained to estimate a complex-valued mask that extracts $\mathbf{s}_m(\tau)$ from $\mathbf{x}_m(\tau)$ and is optimized using the \ac{SNR} loss function \cite{SNRLoss}. This network had approximately $0.5$~M parameters. {When applied across $M$ microphones, this approach provides one source signal estimate per microphone signal and therefore, its results were averaged over the $M$ channels}. 
    
    \item A \ac{DNN}-driven \ac{MVDR} beamformer (denoted \ac{DNN}-\ac{MVDR}) which uses free-field steering vectors steered towards the true source \ac{DOA}. The noise spatial covariance matrices are {recursively estimated} using the estimated noise microphone signals. The noise signals are estimated using complex-valued masks estimated by a pre-trained $\mathbb{C}$RUnet. This approach is used as a representative of \ac{DNN}-supported beamforming methods.
    
    \item An oracle knowledge \ac{MVDR} beamformer (denoted O\ac{MVDR}) which, similarly to the \ac{DNN}-\ac{MVDR}, uses free-field steering vectors steered towards the true source \ac{DOA}. The noise spatial covariance matrices are {recursively estimated} using the ground truth noise microphone signals. This beamformer represents an upper bound for similar methods which rely on estimating the noise components in calculating the spatial covariance matrices. 
    
    \item An oracle knowledge \ac{GMVDR} beamformer (denoted O\ac{GMVDR}) which uses the true \ac{RTF}s calculated w.r.t. the source position in addition to the true noise microphone signals for {recursively estimating} the spatial covariance matrices. This beamformer represents an upper bound for achievable performance using \ac{MVDR} beamformers as it uses oracle spectral and spatial knowledge. 
\end{itemize}

\begin{table}[tbp]	%\vspace*{-.2cm}
	\caption{Average performance of the various approaches.}
	\vspace*{-0.25cm}
	\setlength{\tabcolsep}{5pt}
	\begin{center}
    	    \begin{tabular}{c  c c c c }
        		\toprule  
        		                            & $\Delta$SINR [dB]  & SDR \phantom{ }[dB] & $\Delta$PESQ & $\Delta$STOI   \\
        		\midrule
        		$\mathbb{C}$RUnet            & 7.7     & 4.2        &  0.16      & 0.03  \\ 
        		\ac{DNN}-\ac{MVDR}           & 5.3     & -          &  0.08      & 0.07   \\
        		O\ac{MVDR}                   & 5.0     & -          &  0.1       & 0.09  \\
        		O\ac{GMVDR}                  & 14.3    & -          &  0.24      & 0.12  \\
        		\ac{COSPA}                   & 7.5     & 5.3        &  0.23      & 0.09  \\
        		\bottomrule 
        	\end{tabular} 
		\vspace{-.4cm}
	\end{center}
	\label{tab:tabPerfMeas}
	\vspace{-.3cm}
\end{table}

For all considered algorithms, online processing was carried out for {a linear array with $M=5$ omnidirectional microphones with uniform spacing of $4$~cm, using signal frames} of length $1024$ samples and with frame shifts of $512$ samples for a sampling frequency {of} $f_s=16$~kHz. The \ac{COSPA} was configured with $\{L_1= 260, L_2=L_3=128, L_4=513, L_5=L_6=256\}$ resulting in approximately $2.7$~M free parameters. 

For this evaluation, two datasets were generated. In all datasets, each scenario included one desired speech source and two interferers, a noise source and a music source. The speech utterances were taken from the TIMIT dataset \cite{TIMIT} with disjoint speakers for training and testing. The noise and music sequences were obtained from the MUSAN dataset \cite{musan2015}, {which includes singing voices} among other types of noise, {and for which} training and testing sequences were also disjoint. To generate the training dataset, $6000$ scenarios, each 7~s long, (11hrs 40min) were created. Each scenario consisted of a room of random dimensions between $[3, 3, 1]$~m and {$[8, 8, 4]$}~m and a reverberation time sampled randomly from the range $[0.3-0.7]$~s. The positions of the {microphone array},  desired speech source, noise source and music source were also sampled randomly within the simulated room. The \ac{RIR} of the simulated sources were generated using the image-source method \cite{EHabets}. 

As for the test dataset, 300 scenarios were generated using randomly sampled room dimensions, reverberation times, array, speech source, noise source and music source positions similar to the training dataset. The {inter-microphone distance} was identical across all scenarios in both datasets.

For both datasets, the \ac{SNR} and signal-to-music ratio was sampled randomly per scenario from the range {$[-7, 0]$}~dB, {individually}. In addition, to simulate microphone noise, white additive noise for an \ac{SNR} of 30~dB was added to each microphone signal. 

To compare the different approaches, four different measures are used\footnote{Audio examples and source code implementation can also be found at \url{https://github.com/ModarHalimeh/COSPA}}, averaged over time and scenarios:
\begin{itemize}
    \item $\Delta${SINR}: describes the gain in terms of \ac{SINR} when comparing the \ac{SINR} at the first microphone to that of the enhanced signal. The \ac{SINR} is calculated as the ratio between the energy of the (filtered) source signal to the energy of the (filtered) music and noise signals.
    \item {SDR}: describes the \ac{SDR} as calculated for the (filtered) source signal to quantify the distortions introduced by the filtering \cite{bsseval}.
    \item $\Delta$PESQ: describes the {PESQ} (Perceptual Evaluation of Speech Quality \cite{pesq}) difference between the unprocessed first microphone signal and the enhanced signal.
    \item $\Delta$STOI: describes the {STOI} (Short-Time Objective Intelligibility \cite{stoi}) difference between the unprocessed first microphone signal and the enhanced signal.
\end{itemize}
As a reference signal for the SDR, PESQ and STOI calculations, the \textit{dry} non-reverberant source signal was used.

The averaged performance results are provided in TABLE~\ref{tab:tabPerfMeas}, where the OGMVDR beamformer performs best as it utilizes perfect spatial and spectral knowledge. When comparing the \ac{COSPA} to single-channel $\mathbb{C}$RUnet, clear gains are observed due to the utilization of spatiospectral filtering in comparison to spectral filtering only. We must point out that due to the random nature of the testing dataset, it included scenarios of limited spatial diversity, in which the advantages of spatial filtering are less pronounced, driving the average results of the single-channel approach closer to other multichannel ones. A comparison between OMVDR, OGMVDR and \ac{COSPA} places \ac{COSPA} in terms of performance in-between the two oracle knowledge methods which is very encouraging given that \ac{COSPA} is not provided any side information such as source \ac{DOA}. It is worth mentioning that no \ac{SDR} values are provided for the different \ac{MVDR} beamformer variants, as distortionless response is guaranteed in the source's direction. 

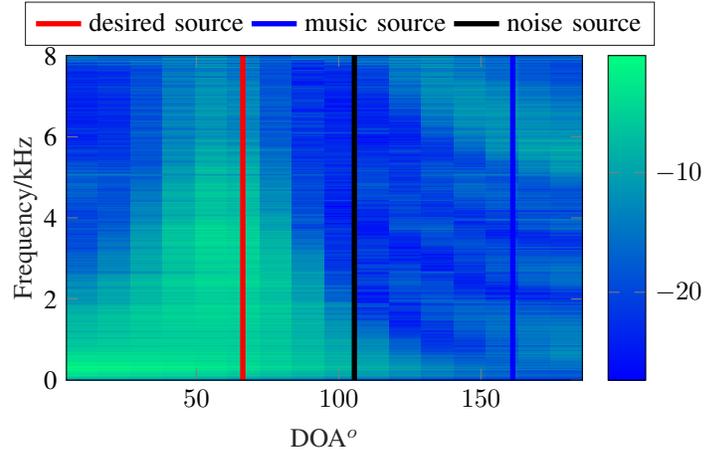
\begin{figure}[tp]
    \centering
\begin{tikzpicture}

\begin{axis}[%
width=2.7in,
height=1.7in,
scale only axis,
point meta min=-27.350513458252,
point meta max=-0.197758615016937,
axis on top,
xmin=4.33333333333333,
xmax=185.666666666667,
xlabel style={font=\color{white!15!black}},
xlabel={DOA$^{o}$},
y dir=reverse,
ymin=-4.02217995472837,
ymax=8000.11592995473,
ylabel style={font=\color{white!15!black}},
yticklabels={8, 8, 6, 4, 2, 0},
ylabel={Frequency/kHz},
y label style={at={(0.1,.5)}},
axis background/.style={fill=white},
legend style={at={(0.53,1.03)}, legend columns=3, anchor=south, legend cell align=left, align=left, draw=white!15!black},
colormap={mymap}{[1pt] rgb(0pt)=(0,0,1); rgb(255pt)=(0,1,0.5)},
colorbar style={at={(1.05,1)}}, 
colorbar
]
\addplot [plot graphics/node/.append style={yscale=-1,anchor=north west}, forget plot] graphics [xmin=4.33333333333333, xmax=185.666666666667, ymin=-4.02217995472837, ymax=8000.11592995473] {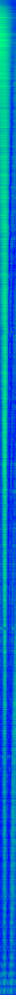};
\addplot [color=red, line width=2.0pt]
  table[row sep=crcr]{%
66.3188420720662	0\\
66.3188420720662	8000\\
};
\addlegendentry{desired source}

\addplot [color=blue, line width=2.0pt]
  table[row sep=crcr]{%
161.223134241789	0\\
161.223134241789	8000\\
};
\addlegendentry{music source}

\addplot [color=black, line width=2.0pt]
  table[row sep=crcr]{%
105.535077225374	0\\
105.535077225374	8000\\
};
\addlegendentry{noise source}

\end{axis}

\end{tikzpicture}%
    \vspace*{-5mm}\caption{An examplary \ac{COSPA} beampattern.}\vspace*{-6mm}
    \label{fig:BP_sample15} 
\end{figure}

To better examine the spatial selectivity of the proposed approach, the average results in Table~\ref{tab:tabPerfMeas} are complemented by the beampattern depicted in Fig.~\ref{fig:BP_sample15}. This beampattern is generated by simulating $36$ equidistant white noise sources placed at different \ac{DOA}s with angular distance increments of $5^{\circ}$, under the free-field propagation assumption. Then, a set of complex-valued masks $\{ \mathbf{\mathcal{M}}_m(\tau); \quad \tau = 1, 2,...\}_{m=1}^M$ is generated for one sample in the test set, i.e., to extract one desired speech signal from a noisy mixture, as described earlier. Using the masks $\{ \mathbf{\mathcal{M}}_m(\tau); \quad \tau = 1, 2,...\}_{m=1}^M$, the white noise sources' microphone signals are filtered, and the power of the filtered signals, averaged over the signals' duration, is depicted in Fig.~\ref{fig:BP_sample15} in [dB] after being normalized to a maximum of $0$~dB. As shown by the beampattern, the proposed \ac{COSPA} is able to successfully localize the desired source as well as being spatially selective to emphasize signals coming from the source's direction. One must point out that since Fig.~\ref{fig:BP_sample15} is averaged over time, an unseen aspect is the time-varying nature of the produced masks that, e.g., can exploit the different sources' activity patterns. Finally, it is worth noting that unlike \ac{MVDR}-based approaches, the \ac{COSPA} does not guarantee a distortionless response in the source's direction which can be seen as a result of the spectral filtering side of the method.

%%%%%%%%%%%%%%%% Concolusion %%%%%%%%%%%%%%%%%%%%%%
 \vspace*{-2mm} \section{Conclusion and Final Remarks} %\vspace*{-1mm}
In this paper we introduced a novel data-driven approach to multichannel signal enhancement. This approach utilizes a complex-valued \ac{DNN}, termed Complex-valued Spatial Autoencoder, to estimate complex-valued masks that are applied to the microphone signals. The proposed approach is compared to different single and multichannel approaches under different acoustic conditions, where the \ac{COSPA}'s spatiospectral filtering capabilities reflect physically plausible spatial selectivity and result in superior speech quality. Finally, encouraged by the results achieved in denoising, we plan on extending \ac{COSPA} to the task of source extraction, where multiple interfering speakers are considered.

\clearpage
	
\bibliographystyle{IEEEbib}
\bibliography{ms}
\end{document}